%% file: output.tex
\documentclass{article}

\usepackage{arxiv}

\usepackage[utf8]{inputenc} 
\usepackage[T1]{fontenc}    
\usepackage{hyperref}       
\usepackage{hyperxmp}
\usepackage{url}            
\usepackage{booktabs}       
\usepackage{amsfonts, amsmath, amsthm, amssymb}       
\usepackage{nicefrac}       
\usepackage{microtype}      
\usepackage{cleveref}       
\usepackage{lipsum}         
\usepackage{graphicx}
\usepackage{biblatex}
\usepackage{doi}
\usepackage{appendix}

\usepackage{tikz}
\usetikzlibrary{positioning, calc, matrix, fit, backgrounds, arrows.meta}
\newcommand\numberthis{\addtocounter{equation}{1}\tag{\theequation}}

\providecommand{\keywords}[1]{\textbf{\textit{Keywords---}} #1}

\bibliography{ref}

\theoremstyle{remark}
\newtheorem{assumption}{Assumption}
\newtheorem{result}{Result}

\hypersetup{
pdftitle={Clarifying the role of placebo response classification in the analysis of the Sequential Parallel Comparison Design},
pdfauthor={Benjamin Stockton, Michele Santacatterina, Soutrik Mandal, Charles M. Cleland, Erinn M. Hade, Nicholas Illenberger, Sharon Meropol, Andrea B. Troxel, Eva Petkova, Chang Yu, and Thaddeus Tarpey},
pdfkeywords={clinical trials, latent variables, placebo response, estimand},
}

\title{Clarifying the role of placebo response classification in the analysis of the Sequential Parallel Comparison Design}

\newcommand{\runninghead}{A Preprint }

\usepackage{authblk}

\author[1]{Benjamin Stockton} 
\author[1]{Michele Santacatterina} 
\author[1]{Soutrik Mandal} 
\author[1]{Charles M. Cleland} 
\author[1]{Erinn M. Hade} 
\author[1]{Nicholas Illenberger} 
\author[1]{Sharon Meropol} 
\author[1]{Andrea B. Troxel} 
\author[1]{Eva Petkova} 
\author[1]{Chang Yu} 
\author[1]{Thaddeus Tarpey}

\affil[1]{Division of Biostatistics, Department of Population Health\newline
New York University Grossman School of Medicine\newline
New York, New York}

\date{\today}

\begin{document}

\maketitle

\begin{abstract}
Sequential parallel comparison design (SPCD) clinical trials aim to
adjust active treatment effect estimates for placebo response to minimize the impact of placebo responders on the estimates. This is
potentially accomplished using a two stage design by measuring treatment
effects among all participants during the first stage, then classifying some
placebo arm participants as placebo non-responders who will be
re-randomized in the second stage. In this paper, we use causal inference tools to clarify under what assumptions treatment effects can be identified in SPCD trials and what effects the conventional estimators target at each stage of the SPCD trial. We further illustrate the highly influential impact of placebo response
misclassification on the second stage estimate. We conclude that the
conventional SPCD estimators do not target meaningful treatment effects. 
\end{abstract}

\keywords{clinical trials, latent variables, placebo response, estimand}

\renewcommand\thefootnote{}
\footnotetext{\textbf{Abbreviations:} SPCD, Sequential Parallel Comparisons Design; ATE, Average Treatment Effect; NR, Non-Responder; PR, Placebo Responder.}

\renewcommand\thefootnote{\fnsymbol{footnote}}
\setcounter{footnote}{1}


\section{Introduction}\label{sec-intro}

Placebo response complicates statistical inference and clinical
understanding of treatment efficacy in 
placebo-controlled clinical trials in psychiatry and other fields\cite{fava2003, chen2011, rybin2018}. As a result, studies have
less ability than expected to detect effective treatments due to estimating a smaller
active treatment effect. Estimating placebo
response is one way to correct for this bias in statistical
inference of average active treatment effects. The sequential parallel
comparison design (SPCD) was developed to address this gap by attempting to identify the placebo non-responders during the first stage of the design, and conducting the second stage only among those identified as non-responders\cite{fava2003}.

\begin{figure}

\centering{

\input{spcd_diagram.tex}

}

\caption{\label{fig-spcd-design}A diagram of the SPCD design. At stage
1, \(N\) participants are randomized to placebo \((A_1 = 0)\) or active
treatment \((A_1=1).\) All participants are either latent true placebo
responders \((L=1)\) or latent true placebo non-responders \((L=0),\) so
each group is a mixture of placebo responders and non-responders. The
placebo-treated participants are classified as placebo responders
\((R=1)\) if their outcome is beyond a specified threshold and otherwise
as placebo non-responders \((R=0).\) At stage 2, the placebo
non-responders are randomized again and assigned to placebo \((A_2=0)\)
or active treatment \((A_2=1).\) Classified placebo responders and those
on active treatment from stage 1 continue on the originally assigned
treatment without re-randomization.}

\end{figure}
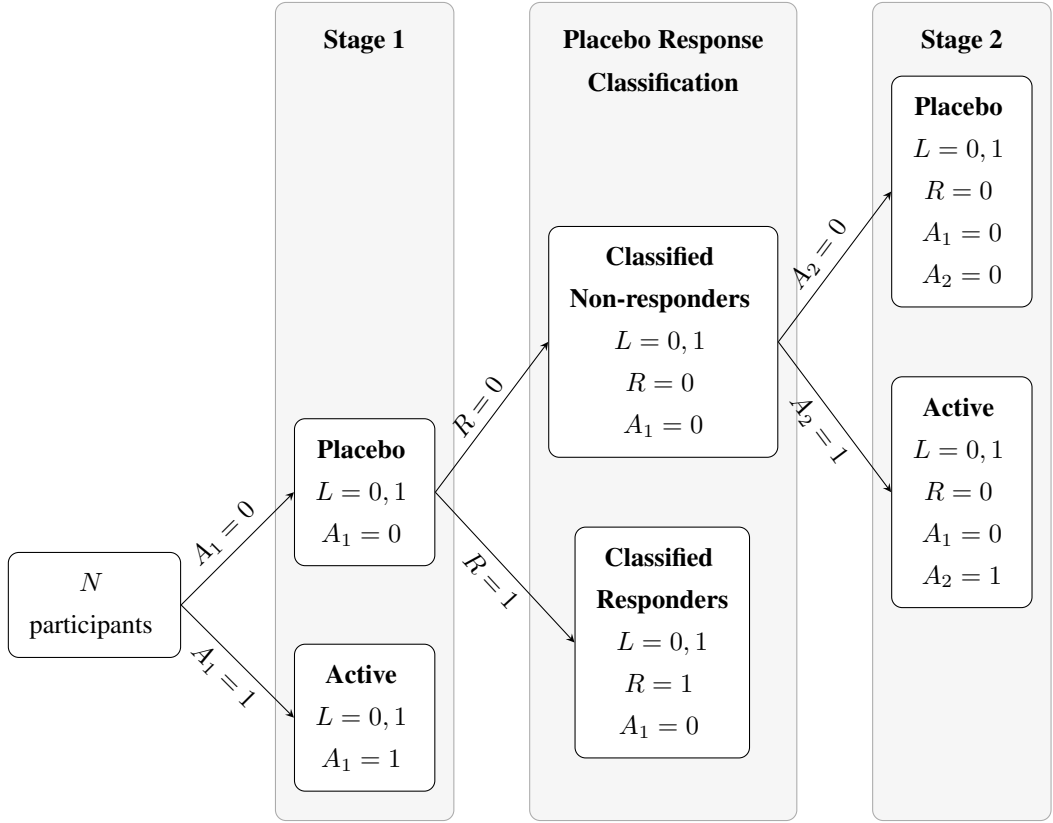

Typically participants are allocated to active and placebo arms at stage 1 using a 2:1 ratio with 2 participants assigned to placebo for each assigned to active treatment. A placebo response classification rule based on symptom improvement is then used to classify the placebo non-responders. 
Those classified as non-responders after stage 1 are then randomized in stage 2 in equal ratio to active treatment or placebo. In total this often results in roughly a third of the participants receiving the active treatment at stage 1, a third receiving placebo and being categorized as placebo responders, and the final third being categorized as placebo non-responders and being re-randomized at stage 2. While alternative strategies to setting randomization allocation ratios are possible\cite{tamura2011}, the decision to choose a particular allocation is typically arbitrary. Additionally, the allocation scheme does not affect our results regarding the estimands and estimators in Section~\ref{sec-estimands}. Figure \ref{fig-spcd-design} displays the
trial design. 
Note that in this paper we use the term estimand, treatment effect, causal effect, and causal parameter interchangeably.


The intuition behind the design is that the placebo response in stage 1 dilutes the measured active treatment effect while the effect measured at stage 2
should only capture the active treatment effect without placebo response. The literature
on statistical analysis for the SPCD proposes an overall estimate
resulting from a weighted average of the stage 1 and stage 2 active treatment
effect estimates\cite{chen2011, rybin2018, tamura2011, chi2015, cui2019}.
However, statistical inference under this design comes with significant
complications\cite{rybin2018, tamura2011, chi2015,
rybin2015, liu2022}. In the rest of the manuscript, we use responders and
non-responders to refer specifically to placebo response, unless
otherwise noted. For brevity, we also use treatment effect or average treatment effect to refer specifically to the active treatment effect or average active treatment effect respectively.

Prior work\cite{chi2015} discussed six issues 
found with the conventional SPCD estimators. These issues include (1)
placebo response biasing estimates, (2) assuming the active treatment effect is
constant over time, (3) uncertainty in how to set the weight
for the overall active treatment effect estimate averaging the stage 1 and stage 2 estimates,
(4) unclear interpretation of the overall estimate, (5) inconsistency between the
directions of the stage 1 and the overall estimate, and (6) the common
assumption that stage 1 and stage 2 are independent.

In this manuscript, we apply the counterfactual and causal inference
framework\cite{rubin1974} to illustrate the
incoherence of the conventional weighted SPCD estimator, and to
highlight an additional complication: placebo responder status
misclassification. Others have discussed misclassification, primarily to
motivate better predictive models\cite{rybin2018, rybin2015,
liu2022}. Our approach to developing estimands and estimators using causal inference tools, discussed in Section~\ref{sec-estimands}, is similar to, but distinct and independently developed from those proposed in recent work by Liu et al.\cite{liu2022} and by Song et al.\cite{song2025}.

In particular, the recent work by Song et al.\cite{song2025} proposes a formulation in which placebo responder status is an unobserved \emph{baseline} trait, and the usual SPCD responder/non-responder classification based on the stage 1 outcome is treated as an imperfect surrogate. Their causal estimands for stage~1 are identified under assumptions that are coherent in this latent-baseline framework in which baseline covariates fully explain the association between responder status and outcomes, and that the probability of being a responder given baseline covariates can be correctly modeled. However, these assumptions do not hold for the SPCD as actually implemented. In practice, responder status is defined only \textit{after} observing the stage 1 outcome. In contrast, we adopt a structural causal model that respects the operational SPCD
classification rule, and we show that under this model these and related
treatment effects are not nonparametrically identified without
additional mixture assumptions, and that the conventional weighted SPCD
estimator does not target a well-defined causal estimand. We provide additional discussion of this
point in our analysis of the ADAPT-A trial in Section~\ref{sec-example},
where we argue that the latent-baseline formulation requires
implausibly strong assumptions and reflects a clear mismatch with the
SPCD design as actually implemented. In addition, we show how crucial
accurate placebo response classification is to accurately estimating the active treatment effects in the SPCD.

In Section~\ref{sec-estimands}, we develop quantities of interest, discuss identification assumptions required for estimation, and derive expectations for the conventional estimators in the SPCD using the causal inference framework. Using these derivations, we argue that the SPCD-specific estimands are not clearly defined and are either not targeted by the conventional estimators due to placebo response misclassification, or are targeted under unrealistic scenarios. Some of the unrealistic scenarios defeat the purpose of using an SPCD design (we provide more details on Sections \ref{stage-2-estimator-non-responders}, \ref{overall-weighted-estimator}, and \ref{sec-discussion}). 
The stage 1 estimand and estimator do target the average treatment effect for all participants, however this is achievable with a standard parallel randomized design. We demonstrate the importance of accurate placebo response classification using a simulation study in Section~\ref{sec-simulation}
and with an illustration using a depression trial example in Section~\ref{sec-example}.
Finally, in Section~\ref{sec-discussion}, we summarize our results, connect them to previously published work, and provide some direction on how to approach analysis for SPCD trials.



    
    


\section{Background}\label{sec-background}

In this section, we introduce some common notation to clarify the
meanings of various estimands and estimators previously proposed in the
literature.

\subsection{SPCD Set Up}\label{sec-spcd-setup}

We consider the binary latent placebo responder framework proposed by
Rybin et al.\cite{rybin2018} wherein the population is
a mixture of placebo responders and non-responders. A variation of the
latent responder status is also used more generally as a counterfactual to categorize both drug
response and placebo response\cite{liu2022}. Both  latent responder approaches\cite{rybin2018, liu2022} use an expectation maximization (EM)
algorithm\cite{dempster1977} to account for the unknown responder status when fitting their
respective models.

\subsubsection{Notation}

For each of $i\in\{1,\ldots,n\}$ study participants, let \(L_i\)
denote a binary latent variable indicating whether a participant would respond to placebo (\(L_i = 1\)) or not (\(L_i = 0\)). Let  \(Y_{i,0}\), \(Y_{i,1}\), and \(Y_{i,2}\) denote the participant's baseline (e.g., pre-treatment symptom severity), stage 1 and stage 2 outcomes, respectively. Let \(A_{i,1}\) denote the randomized arm at Stage 1, i.e, placebo (\(A_{i,1} = 0\)) or active treatment (\(A_{i,1} = 1\)) and \(A_{i,2}\) the assigned or randomized arm at stage 2, respectively. Finally, let $R_i$ represent the placebo-responder indicator, i.e., participants are \emph{classified} as responders (\(R_i = 1\)) or non-responders (\(R_i = 0\)). The observed data are $D=(Z_1,\ldots,Z_n)$, where $Z_i$ represents the data for the $i^{th}$ participant, i.e.,
$Z_i=(Y_{i,0}, Y_{i,1}, Y_{i,2}, R_i, A_{i,1}, A_{i,2})$ following some multivariate probability distribution. The latent exogenous variables $U_{Y_0}$, $U_{A_1}$, $U_{Y_1}$, and $U_{Y_2}$ are unmeasured and contain background information or influences on the observed variables\cite{pearl2009causality}.


\subsubsection{A structural causal model and associated DAG}

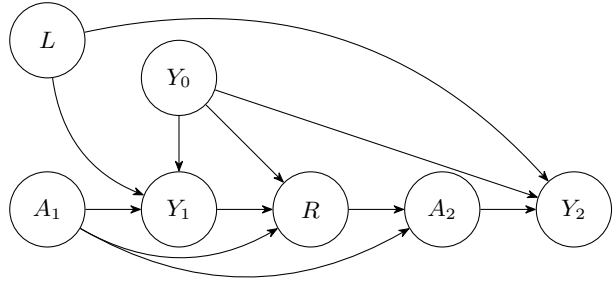
\begin{figure}
\begin{minipage}{0.35\textwidth}
\centering
\begin{align}
Y_{i,0} &= g_{Y_0}(U_{i,Y_0}),\notag\\
A_{i,1} &= g_{A_1}(U_{i,{A_1}}),\notag\\
Y_{i,1} &= g_{Y_1}(Y_{i,0}, L_i, A_{i,1}, U_{i,Y_1}),\notag\\
R_i   &= C_R(Y_{i,0}, Y_{i,1}),\notag\\
A_{i,2} &= 
\begin{cases}
g_{A_2}(U_{i,A_2}) &\text{if } A_{i,1}=0,\;R_i=0,\\
0 &\text{if } A_{i,1}=0,\;R_i=1,\\
1 &\text{if } A_{i,1}=1,
\end{cases}\notag\\
Y_{i,2} &= g_{Y_2}(Y_{i,0}, L_i, A_{i,1}, A_{i,2}, U_{i,Y_2}).\notag
\end{align}
\end{minipage}%
\hspace{6mm}
\begin{minipage}{0.55\textwidth}
\small
\centering
\begin{tikzpicture}[
  node distance=1.75cm,
  every node/.style={circle, draw, minimum size=1cm},
  >={Stealth[round]}
]
\node (A1) {$A_1$};
\node (L) [above of=A1, yshift=5mm] {$L$};
\node (Y0) [right of=L, yshift=-5mm] {$Y_0$};
\node (Y1) [right of=A1] {$Y_1$};
\node (R)  [right of=Y1] {$R$};
\node (A2) [right of=R] {$A_2$};
\node (Y2) [right of=A2] {$Y_2$};

\draw[->,bend right] (L) to (Y1);
\draw[->,bend left]  (L) to (Y2);
\draw[->] (Y0) -- (Y1);
\draw[->] (Y0) -- (Y2);
\draw[->] (Y0) -- (R);
\draw[->] (A1) -- (Y1);
\draw[->,bend right]  (A1) to (R);
\draw[->,bend right]  (A1) to (A2);
\draw[->] (Y1) -- (R);
\draw[->] (R)  -- (A2);
\draw[->] (A2) -- (Y2);
\end{tikzpicture}
\end{minipage}%
\caption{\label{fig-scm}This causal DAG depicts the structural causal
model underlying the SPCD with the addition of our latent placebo
responder status \(L.\)}
\end{figure}

To clarify the SPCD set up, we use the structural causal model (SCM) and directed acyclic graph (DAG)\cite{pearl2009causality}  described in Figure~\ref{fig-scm}. From the next paragraph, we omit the participant index $i$ for better readability. 

In this SCM, $L$, $U_{Y_0}$, $U_{A_1}$, $U_{Y_1}$, and $U_{Y_2}$ are latent (exogenous) variables. The baseline symptom severity $Y_0$ depends only on exogenous variables $U_{L}$, and $U_{Y_0}$. In Stage 1, participants are randomized to placebo or active treatment following the randomization mechanism $g_{A_1}(\cdot)$. The Stage 1 outcome \(Y_1\) depends on baseline symptom severity ($Y_0$), latent (true) placebo response status ($L$), treatment assignment ($A_1$), and exogenous factors. Placebo-treated participants are \emph{classified} as responders (\(R = 1\)) or non-responders (\(R = 0\)) using a classifier \(R = C_R(Y_0, Y_1)\) based on baseline and stage 1 data \((Y_0, Y_1).\) In Stage 2, placebo-treated participants \emph{classified} as placebo non-responders are re-randomized following $f_{A_2}$ to placebo or active treatment, while all others continue with their Stage 1 treatment, (\(A_2\)). The final outcome \(Y_2\) depends on baseline severity ($Y_0$), latent response type ($L$), and both treatment assignments ($A_1,A_2$). $Y_2$ does not depend on $R$ conditional on $A_2$ because $R$ only determines whether stage 2 treatment ($A_2$) will be re-randomized (when $R = 0$ and $A_1=0$), otherwise $A_2 = A_1$. The stage 2 treatment ($A_2$) and latent placebo response status ($L$) then contain all the influence of $R$ on $Y_2$. 



It is important to note that, while \(L\) is unobserved, it is
conceptually treated as a baseline variable in terms of causal
temporality. It represents an inherent placebo response status that
exists prior to treatment assignment and affects post-baseline outcomes,
including \(Y_1\) and \(Y_2\). In contrast, the observed placebo
response classification \(R\) is not a baseline variable---it is
deterministically defined as a function of \(Y_0\) and \(Y_1\), and thus
occurs \emph{after} the outcome \(Y_1\) is realized. From a causal
ordering perspective, this places \(L\) upstream of all outcomes, while
\(R\) is downstream of \(Y_1\).

To put the latent and classified responders in context, consider pain
score as an outcome. Conceptually, we consider a true placebo responder
\((L=1)\) to be a participant for whom the placebo treatment provides some therapeutic benefit compared to if they had not been treated at
all. A true placebo non-responder \((L=0)\) is someone whose potential pain score
trajectory is unaffected by receiving the placebo.
However, due to variance in the outcomes, it is plausible that a true
placebo responder \((L=1)\) 
may be misclassified as a non-responder
\((R=0)\) due to insufficient improvement to meet the responder threshold
and vice versa for true placebo non-responders. As a result, we would
very likely still have a mixture of true placebo responders and
non-responders in the second stage of the SPCD, which would limit the
ability to estimate the average active treatment effect without any placebo
response.

\section{SPCD Estimands and Estimators}\label{sec-estimands}

We discuss potential estimands targeted in a SPCD in
Section~\ref{what-are-the-estimands} as functions of counterfactual variables\cite{pearl2009causality}. 
Then, we introduce assumptions for their identification in
Section~\ref{sec-causal-assumptions}. 
Finally, we discuss conventional SPCD estimators  in
Section~\ref{sec-prev-estimators}.

We assume all population average active treatment effects are time
invariant, meaning the effect of the active treatment compared to
placebo is the same at each stage for each group: the whole population
(mixture of placebo responders and non-responders), the true placebo
non-responders, and the true placebo responders. Let \(\Delta_{\mathrm{all}}\) be the
average treatment effect in the population as a whole (Equation
\ref{eq-ate-all-2}), \(\Delta_{NR}\) be the average treatment effect for
true placebo non-responders (Equation~\ref{eq-ate-nr-2}),
\(\Delta_{PR}\) be the average treatment effect for true placebo
responders (Equation~\ref{eq-ate-r}), and \(\Delta_{\mathrm{placebo}}\)
be the average effect due to placebo response among placebo responders
relative to the non-responders (Equation~\ref{eq-ate-placebo-2}). Each
of these will be defined and discussed in further detail in
Section~\ref{what-are-the-estimands}.

The stage 1 SPCD estimator is denoted \(\hat{\theta}_1\) (Equation
\ref{eq-theta-1}) and the stage 2 SPCD estimator for participants
classified as placebo non-responders is \(\hat{\theta}_2\) (Equation
\ref{eq-theta-2}). The overall weighted estimator is a weighted average
of the stage 1 and stage 2 estimators with pre-specified weight \(0 \leq w \leq 1\);
\(\hat{\theta}_w = w\hat{\theta}_1 + (1-w)\hat{\theta}_2\) (Equation
\ref{eq-theta-w}). All three estimators are defined in detail in
Section~\ref{sec-prev-estimators}.

\subsection{SPCD Estimands}\label{what-are-the-estimands}


Denote the counterfactual outcomes at times \(t=1,2\) as
\(Y_t^{a,l}\) with the superscript treatment assignment \(a\) and latent
placebo response status \(l\). The
latent placebo response is included in the counterfactual notation to
emphasize that this is an unobserved property of the counterfactual outcome.
Let \(\mu_t^{a,l} = E(Y_t^{a,l})\) be the
expected value at stage $t$ of the outcome after treatment \(a\) with latent placebo
status \(l\), and \(p_L = P(L = 1)\) be the prevalence of true placebo responders in the population.

Let us also assume that \(Y_0\) follows a single baseline distribution without any effects due
to treatment or placebo response by randomization. Then
\(E(Y_0 | A_1, A_2) = E(Y_0) = \mu_0.\)

Four estimands are summarized in Table~\ref{tbl-ate} in terms of average effects (e.g., Average Treatment Effect, ATE): the
average placebo response effect \(\Delta_{\mathrm{placebo}}\) given
by Equation~\ref{eq-ate-placebo-2}, the ATE for placebo non-responders
\(\Delta_{NR}\) by Equation~\ref{eq-ate-nr-2}, the ATE for placebo
responders \(\Delta_{PR}\) by Equation~\ref{eq-ate-r}, and the ATE for
all \(\Delta_{\mathrm{all}}\) by Equation~\ref{eq-ate-all-2}. 

\begin{equation}
	\Delta_{\mathrm{placebo}} = E(Y_1^{a_1 = 0, l=1}) - E(Y_1^{a_1 = 0, l=0})
\label{eq-ate-placebo-2}
\end{equation}

\begin{equation}
	\Delta_{NR} = E(Y_1^{a_1=1}) - E(Y_1^{a_1=0, l=0})
\label{eq-ate-nr-2}
\end{equation}

\begin{equation}
	\Delta_{PR} = \Delta_{NR} - \Delta_{\mathrm{placebo}}
\label{eq-ate-r}
\end{equation}

\begin{equation}
	\Delta_{\mathrm{all} } = \Delta_{NR} - p_L \Delta_{\mathrm{placebo}}
\label{eq-ate-all-2}
\end{equation}

Most previous literature\cite{chen2011, rybin2018, tamura2011, chi2015, cui2019, rybin2015} defines the
average treatment effect for all participants \(\Delta_{\mathrm{all}}\) as
the effect between treated participants and a mixture of placebo
responders and non-responders. The treatment
effects \(\Delta_{NR}\) for the placebo non-responders, and
\(\Delta_{PR}\) for placebo responders are defined less often in the literature\cite{rybin2018, chi2015}. 
We consider \(\Delta_{PR}\) to be
equivalent to \(\Delta_{NR} - \Delta_{\mathrm{placebo}}.\) To our
knowledge, previous work has not defined a quantity equivalent to
\(\Delta_{\mathrm{placebo}},\) the size of the placebo response for
placebo responders relative to placebo non-responders.

Table~\ref{tbl-ate} clarifies two important points:  1) estimands targeted by SPCD depend on unobserved non-responder and/or responder outcomes \(Y_t^{a=0, l = 0}\) and \(Y_t^{a=0,l=1}\), for which clear and transparent assumptions are required \footnote{Note how the placebo treated outcomes \(Y_t^{a=0, l}\) can be expressed as a
mixture between the unobserved non-responder outcome \(Y_t^{a=0, l = 0}\)
and unobserved responder outcome \(Y_t^{a=0,l=1}\), $Y_t^{a = 0, l} = I(L = 1) Y_t^{a = 0, l=1} + (1 - I(L = 1))Y_t^{a = 0, l = 0}$.}, and 2) \(\Delta_{\mathrm{all}}\) and \(\Delta_{NR}\) would be equivalent if the magnitude of any placebo response were zero \((\Delta_{\mathrm{placebo}} = 0\)) or if there were no placebo
responders (\(p_L = 0\)).  If either of these conditions were true, then the rationale for using the SPCD would vanish.

\begin{center}
\begin{table*}[!h]%
\caption{Average effects as functions of counterfactual outcomes and latent
placebo responder status. Average treatment effect is abbreviated as
ATE.}\label{tbl-ate}
\begin{tabular*}{\textwidth}{@{\extracolsep\fill}lcr@{}}
\toprule
\textbf{Quantity} & \textbf{Notation}  & \textbf{Population Subset}    \\
\midrule
Average Placebo Response Effect &
$\Delta_{\mathrm{placebo}} = E(Y_1^{a_1 = 0, l=1}) - E(Y_1^{a_1 = 0, l=0})$
& Stage 1 Placebo-treated; \(A_t = 0\) \\
ATE for Placebo Non-Responders &
$\Delta_{NR} = E(Y_1^{a_1=1}) - E(Y_1^{a_1=0, l=0})$
& Placebo Non-responders; \(L = 0\) \\
ATE for Placebo Responders &
$\Delta_{PR} = \Delta_{NR} - \Delta_{\mathrm{placebo}}$
& Placebo Responders; \(L = 1\) \\
ATE for All Participants &
$\Delta_{\mathrm{all} } = \Delta_{NR} - p_L \Delta_{\mathrm{placebo}}$
& All Participants \\
\bottomrule
\end{tabular*}
\end{table*}
\end{center}

\normalsize



\subsection{Identification 
Assumptions}\label{sec-causal-assumptions}

We provide below causal and statistical assumptions needed to identify the estimands discussed in our previous section. For brevity, we define these assumptions for the Stage 1 counterfactual outcomes, but they can be naturally extended to Stage 2 as well.

\begin{assumption}[Consistency and the Stable Unit Treatment Values Assumption (SUTVA)] \label{asm-cons_sutva}
$$Y_1 = Y_1^{A_1, L} \quad \text{and} \quad Y_{i,1}^{a,l} \perp\!\!\!\perp A_{j,1} \text{ for all } i \neq j.$$
\end{assumption}

Assumption \ref{asm-cons_sutva} states that each unit's observed outcome
corresponds to their latent placebo response status and their counterfactual outcome under the treatment actually
received and that treatment is consistently applied or given across individuals. In other words, there are not different versions of the placebo or active treatment due to variation in providers, dosage, adherence, etc.
It also assumes no
interference between units. Together these comprise the stable unit treatment values assumption\cite{rubin1980, robins2000, hernan2020}. This is encoded in the structural causal
model (Figure \ref{fig-scm}). 

\begin{assumption}[Ignorability] \label{asm-exch}
$$A_1 \perp\!\!\!\perp (Y_1^{1,1}, Y_1^{1,0}, Y_1^{0,1}, Y_1^{0,0})$$
\end{assumption}

Assumption \ref{asm-exch} is satisfied by proper randomization and holds
by design.

\begin{assumption}[Positivity] \label{asm-positivity}
$$0 < P(A_1 = a) < 1 \quad \text{for all }\; a.$$
\end{assumption}

Assumption \ref{asm-positivity} ensures that all treatment arms are
possible and holds by design.

At Stage 2, only placebo-treated participants classified as non-responders are re-assigned, so we need
\(P(A_{2} = a| R_{1} = 0, A_{1} = 0) > 0\) for classified placebo
non-responders and stage 2 treatment assignments \(a = 0, 1.\) If
re-randomized, then this should also hold by design.

\begin{assumption}[Latent-type stability] \label{asm-latent}
$$L \perp\!\!\!\perp A_1$$
\end{assumption}

Assumption \ref{asm-latent} asserts that the latent placebo response
 \(L\) is fixed at baseline and unaffected by treatment assignment. This can
fail if the drug biologically alters placebo pathways (e.g., via neural
priming), so that \(L^{a_1=1} \neq L^{a_1=0}\). Then the mean outcome
under placebo for placebo responders, \(\mu_t^{a=0,l=1}\), no longer
represents the counterfactual placebo outcome for the drug-treated
group, breaking the validity of
\(\Delta_{\text{placebo}} = \mu_t^{a=0,l=1} - \mu_t^{a=0,l=0}\).

\begin{assumption}[Effect homogeneity across $L$] \label{asm-homogeneity}
\begin{equation}
    \mu_t^{a=1,l=1} = \mu_t^{a=1,l=0}
\end{equation}\label{eq-eff-homogeneity}
\end{assumption}

Assumption \ref{asm-homogeneity} posits that the active treatment effect is the
same regardless of latent placebo response status. 

If $\mu_t^{a=1,l=1} \neq \mu_t^{a=1,l=0},$ then we are unable to identify the ATE for non-responders ($L=0$) because we do not classify active treatment participants as responders/non-responders in order to be able to estimate $\mu_t^{a=1,l=0}.$ While this assumption could be violated in practice, we make it here for brevity in the exposition of the issues with the estimators. Alternative strategies that do not require this assumption have been proposed but are beyond the scope of this paper\cite{liu2022}.

Using the latent placebo response variable, we can express the placebo outcome at stage 1 as the mixture
\(Y_1^{a = 0, l} = I(L = 1) Y_1^{a_1 = 0, l=1} + (1 - I(L = 1))Y_1^{a_1 = 0, l = 0},\) \(I(\cdot)\) is an indicator function.

\begin{assumption}[Mixture identifiability in the placebo arm (model-based)] \label{asm-mixture}

Given that $Y_1$ follows a mixture, let $p_L = P(L=1)$ be the unknown placebo responder prevalence, $f_{Y_1}(\cdot; \xi)$ be component densities of $Y_1^{l}$ belonging to a parametric family $\mathcal{F} = \{f(\cdot; \xi)\}$, and $\xi_l$ be parameters indexing the family $\mathcal{F}$ for latent response status $L = l$. Then the density of the stage 1 outcome in the placebo arm is,
\begin{equation}
    f_{Y_1 \mid A_1=0}(y) = p_L f_{Y_1}(y;\xi_1) + (1 - p_L) f_{Y_1}(y;\xi_0),
\end{equation}\label{eq-mixture} 

\noindent such that the mapping $(p_L, \xi_0, \xi_1) \mapsto f_{Y_1|A=0}(\cdot)$ is one-to-one within that model family, and the mixture is identified under that parametric model.

\end{assumption}

Assumption \ref{asm-mixture} is needed to identify \(\mu_1^{0,l} = E(Y_{1}^{0,l})\) for $l = 0,1$ from the observed Stage-1 placebo data. If
\(\xi_1 \approx \xi_0\) or \(p_L \approx 0\), the mixture model is
weakly identified.

\begin{equation}
E(Y_1\mid A_1=0) \;=\; \int y \, f_{Y_1 \mid A_1=0}(y)\, dy
\;=\; p_L\,\mu_1^{0,1} + (1-p_L)\,\mu_1^{0,0},
\end{equation}\label{eq-mixture-mean}

To illustrate, consider the average placebo effect $\Delta_{\mathrm{placebo}} = \mu_1^{0,1} - \mu_1^{0,0}$ and the latent placebo responder status $L$. Let $f_{L=0}(y)$ and $f_{L=1}(y)$ be nonparametric component densities of the stage 1 placebo outcome under $L = 0$ and $L=1$, respectively. From the data we observe only the placebo-arm mixture:
\begin{equation}
f_{Y_1 \mid A_1=0}(y) \;=\; p_L\, f_{L=1}(y) + (1-p_L)\, f_{L=0}(y).
\end{equation}\label{eq-mixture-np-dens}

\noindent Then, nonparametrically, the triple $(p_L,f_{L=0},f_{L=1})$ is \emph{not} point identified from a single observed mixture $f_{Y_1 \mid A_1=0}$, because many different triples can generate the same mixture. Hence $\mu_1^{0,0}$ and $\mu_1^{0,1}$ (and therefore $\Delta_{\mathrm{placebo}}$) are not 
nonparametrically identified. 

Assumption~\ref{asm-mixture} imposes a parametric mixture structure, as in Equation~\ref{eq-mixture}
which enables \emph{model-based} identification: if the mapping
\(
(p_L,\xi_0,\xi_1)\;\longmapsto\; f_{Y_1 \mid A_1=0}(\cdot)
\)
is one-to-one within the chosen family $\mathcal{F}$, then $(p_L,\xi_0,\xi_1)$ and thus $(\mu_1^{0,0},\mu_1^{0,1})$ are identified \emph{under that model}. Note this identifies component means, but does not imply we can correctly classify individual $L$’s. While other identification strategies for causal inference with latent variables have been proposed (see, e.g., \cite{zhu2024causal,zhang2021bounding}), in this manuscript we adopt a model-based approach. As our goal is to clarify causal effects in SPCD trials, the model-based framework suffices for that purpose, as shown in the following sections and in the Supplementary Materials.

We also make one additional \textit{statistical} assumption of stationarity for the average effects.

\begin{assumption}[Stationarity] \label{asm-stationarity}
$E(\Delta_{g,1}) = E(\Delta_{g,2}) = \Delta_{g}$ and $Var(\Delta_{g,1}) = Var(\Delta_{g,2})$ where $\Delta_{g, t}$ is the average effect for sub-group $g \in \{\mathrm{all}, NR, PR, \mathrm{placebo}\}$ at stage $t = 1,2.$
\end{assumption}

Assumption \ref{asm-stationarity} assumes that the effects remain
constant over time and between stages. This form of stationarity is
necessary for statistical inference based on the conventional SPCD estimators.


Note that 
we do not assume that
\(E(Y_1^{a_1 = 0, l = 0}) = E(Y_0)\) as this assumption is unlikely to
hold, e.g.,~in psychiatric or pain treatment trials where we may expect
conditions to improve over time regardless of treatment or placebo
effects. Both psychiatric and pain treatments are common applications of
the SPCD.

Based on these assumptions, we provide identification of Table~\ref{tbl-ate} estimands in Appendix ~\ref{sec-app-est-deriv} and that will be used to clarify what estimands are targeted by the conventional SPCD estimators.

\subsection{The Conventional SPCD Estimators}\label{sec-prev-estimators}

Let the total \(N\) participants be partitioned into \(N_{A}\)
participants assigned to active treatment in stage 1, and \(N_{P}\)
participants assigned to placebo in stage 1. Let \(N_{NR}\) and
\(N_{PR}\) be the random number of classified placebo non-responders and
responders respectively, \(N_{NR} + N_{PR} = N_{P};\)
\(N_{NR} = \sum_{i=1}^{N_P} I(R = 0)\) for some placebo response
indicator \(R\); and let \(n_{NR}\) and \(n_{PR}\) be the observed
counts of \(N_{NR}\) and \(N_{PR}\) respectively. The number of
non-responders \(N_{NR}\) is further split into \(N_{PP}\) and
\(N_{PA}\) for placebo-placebo assignment and placebo-active assignment
with observed counts \(n_{PP}\) and \(n_{PA}\) respectively, such that
\(N_{PP} + N_{PA} = N_{NR}\) and \(n_{PP} + n_{PA} = n_{NR}.\)

The commonly used weighted estimator\cite{chen2011, rybin2018, tamura2011, chi2015, cui2019,
liu2022} combines stage 1, \(\hat{\theta}_1\),
and stage 2, \(\hat{\theta}_2\), to get \(\hat{\theta}_w\) using an
pre-determined scalar weight \(w\). See
Appendix~\ref{sec-app-theta-hat-deriv} for full derivations of the
expectations.

\subsubsection{Stage 1 Estimator (All
Participants)}\label{stage-1-estimator-all-participants}

The stage 1 estimator \(\hat{\theta}_1\) is defined as in Equation
\ref{eq-theta-1},

\begin{align*}
\hat{\theta}_1 = &\frac{1}{n_A} \sum_{i=1}^{n} (Y_{i, 1} - Y_{i,0})I(A_{i,1} = 1) - \frac{1}{n_P} \sum_{i=1}^{n} (Y_{i, 1} - Y_{i,0})I(A_{i,1} = 0), \label{eq-theta-1} \numberthis 
\end{align*}

\noindent which ideally corresponds to the ATE of the population
\(\Delta_{\mathrm{all}}\) at the end of stage 1.

Under Assumptions \ref{asm-cons_sutva}-\ref{asm-stationarity}, its
expectation is as given in Equation \ref{eq-exp-theta-1}.

\begin{align*}
    E(\hat{\theta}_1) &= \Delta_{NR}- p_L\Delta_{\mathrm{placebo}} \\
    &= \Delta_{\mathrm{all}}. \label{eq-exp-theta-1} \numberthis 
\end{align*}

\noindent Note that we do not condition on \(R\) or \(L\) in this expectation
calculation. The placebo response is marginalized out by the mixture
\(Y_1^{a = 0, l} = I(L = 1) Y_1^{a_1 = 0, l=1} + (1 - I(L = 1))Y_1^{a_1 = 0, l = 0}.\)
So we can expect this estimator to be unbiased for
\(\Delta_{\mathrm{all}}\) in all scenarios.

\subsubsection{Stage 2 Estimator
(Non-Responders)}\label{stage-2-estimator-non-responders}

Given the number of placebo non-responders \(n_{NR} = n_{PP} + n_{PA},\)
the stage 2 estimator can be defined as in Equation \ref{eq-theta-2},

\begin{align*}
\hat{\theta}_2 = &\frac{1}{n_{PA}} \sum_{i=1}^{n} (Y_{i, 2} - Y_{i,1})I(A_{i,1} = 0, R_i = 0, A_{i,2} = 1) \\
&~~~~ - \frac{1}{n_{PP}} \sum_{i=1}^{n} (Y_{i, 2} - Y_{i,1})I(A_{i,1} = 0, R_i = 0, A_{i,2} = 0). \label{eq-theta-2} \numberthis 
\end{align*}

We would like for \(\hat{\theta}_2\) to unbiasedly estimate \(\Delta_{NR}.\) 
Under Assumptions
\ref{asm-cons_sutva}-\ref{asm-stationarity} (modified accordingly for Stage 2),
\(E(\hat{\theta}_2)\) is derived as in Equation \ref{eq-exp-theta-2},

\begin{align*}
E(\hat{\theta}_2) &= \Delta_{NR} - P(L = 1 | R = 0) \Delta_{\mathrm{placebo}} \\
&\neq \Delta_{NR}. \label{eq-exp-theta-2} \numberthis 
\end{align*}

\noindent \(E(\hat{\theta}_2)\) is neither the ATE for all participants
(\(\Delta_{\mathrm{all}}\)) nor the ATE for the placebo non-responders
(\(\Delta_{NR}\)). This is because \(R \neq L,\) so that the population
in stage 2 is still some mix of true responders and non-responders,
which in turn suppresses the estimate of \(\Delta_{NR}\) as we will see
in the simulation. If we want \(E(\hat{\theta}_2) = \Delta_{NR},\) then
we would need \(P(L=1|R=0) = 0\) meaning the classifier
\(C_R(y_0, y_1)\) must have perfect negative predictive power.
Alternatively, if \(P(L=1|R=0) = P(L=1),\) then
\(E(\hat{\theta}_2) = \Delta_{NR} - P(L = 1) \Delta_{\mathrm{placebo}} = \Delta_{\mathrm{all}},\)
but that would imply that \(R\) is independent of \(L.\) (This may
happen if \(C_R(y_0, y_1) = 0\) for all \(y_0, y_1\), i.e.~everyone is
classified as a placebo non-responder.) Note that we cannot evaluate the
predictive capabilities of \(C_R(y_0, y_1)\) in practice because \(L\)
is latent.

\subsubsection{Overall Weighted
Estimator}\label{overall-weighted-estimator}

Finally, the overall estimator \(\hat{\theta}_w\) is a weighted average
of \(\hat{\theta}_1\) and \(\hat{\theta}_2,\) given by
Equation~\ref{eq-theta-w},

\begin{equation}\phantomsection\label{eq-theta-w}{
\hat{\theta}_w = w \hat{\theta}_1 + (1 - w) \hat{\theta}_2.
}\end{equation}

Under Assumptions \ref{asm-cons_sutva}-\ref{asm-stationarity},

\begin{align*}
E(\hat{\theta}_w) &= wE(\hat{\theta}_{1}) + (1 - w)E(\hat{\theta}_2) \\
&= \Delta_{NR} - \Delta_{\mathrm{placebo}}\left[wP(L=1) + (1-w)P(L=1|R=0)\right]. \label{eq-exp-theta-w} \numberthis 
\end{align*}

\noindent Like \(E(\hat{\theta}_2),\) \(E(\hat{\theta}_w)\) is also ill-defined in
that it targets neither \(\Delta_{\mathrm{all}}\) nor
\(\Delta_{NR},\) as well as mixing the marginal and conditional
probabilities of latent placebo response. The estimator
\(\hat{\theta}_w\) will be unbiased if
\(\Delta_{\mathrm{all}} = \Delta_{NR}.\) But this is only true when
there are only placebo non-responders so that the estimands coincide,
e.g.,~when \(\Delta_{\mathrm{placebo}} = 0\) or \(P(L = 1) = 0.\) The
quantities \(\Delta_{\mathrm{placebo}},~~ P(L = 1 | R=0),\) and
\(P(L = 1)\) are unobservable and cannot be directly estimated because
\(L = l\) is latent, so these conditions cannot be checked.

These results hold generally for the conventional SPCD estimators, however we
have not specified two other crucial analysis decisions; (i) how to
choose \(w\) to actually weight the combination, and
(ii) how to choose the classification function \(C_R(y_0, y_1).\)

A similar latent responder status framework could be used for active
treatment response such that participants may or may not respond to the
active treatment. However, active treatment non-responders are then
effectively receiving a placebo and so are part of the placebo responder
status framework as well. This is a key part of the latent responder
status framework\cite{liu2022}.


\subsection{The (Lack of) Attention Given to
Classification}\label{sec-misclass}

A limitation of previous work on SPCD is the absence of substantial
discussion of placebo response classification, with some recent exceptions\cite{rybin2018, rybin2015, liu2022, song2025}. Typically, classification is
done by setting a cut-off or threshold \(c\) as in
Equation~\ref{eq-thresh-classifier},

\begin{equation}\phantomsection\label{eq-thresh-classifier}{
C_R(y_0, y_1) = I(y_1 - y_0 > c) ~~ \mathrm{or} ~~ C_R(y_0, y_1) = I(y_1 > c),
}\end{equation}

\noindent so that participants with
large improvements over baseline or large outcomes in the placebo arm at
Stage 1 are labeled as placebo responders. As previously noted\cite{rybin2018, rybin2015, liu2022}, this approach is subjective due to the
arbitrary choice of \(c\), error-prone, and often not supported by any
relevant clinical data. However, given that the placebo response status
is unobservable, this is a relatively intractable classification
problem\cite{song2025}.

One attempt\cite{rybin2015} at addressing the misclassification issue proposed using a
re-randomization design that re-assigns treatment to all Stage 1 placebo
participants rather than just placebo non-responders and then weights
their response in Stage 2 by a measure of their responder status. Another approach\cite{rybin2018} went further in the discussion of
misclassification and proposed using a normal mixture model to classify
the responders and non-responders.

This mixture model approach was expanded upon by Liu et al. (2022)\cite{liu2022} and the misclassification issue was given more extensive discussion. They note perfect classification as being a
crucial assumption in the derivation of their strata-level effects,
estimators, and inference. Liu et al.\cite{liu2022}
also criticize the use of thresholds as subjective and error-prone which
can potentially bias the inference. Instead they also propose fitting a
mixture model with an EM algorithm and using the fitted
probabilities of belonging to each strata given the data to then sample
the ``observed'' responder statuses \(S_i^{a_1}\) and \(S_i^{a_2}\) for
each individual \(i.\) However, the sampled responder statuses are still
treated as known. They discuss how an approach that treats these
statuses as stochastic may be an avenue for further development.
The stochastic approach is beyond the scope of this paper.
Liu et al.'s simulation studies\cite{liu2022}
demonstrate that the causal estimators derived through their approach are biased when assuming the sampled placebo response are fixed. This is likely due to
the misclassification which they note in their discussion.

\section{Simulation Study}\label{sec-simulation}

In order to illustrate the bias of $\hat{\theta}_2$ and $\hat{\theta}_w$ for $\Delta_{NR}$ and $\Delta_{\mathrm{all}}$ respectively, we devised a simulation study to evaluate the impacts of the size of the placebo response $\Delta_{\mathrm{placebo}}$ (also called the average placebo effect or APE) and error standard deviation $\sigma_\epsilon$ on the misclassification of placebo non-response and the resulting bias in the estimators. 

In these simulations, we assume that the baseline \(Y_0\) is independent
of treatment assignments \(A_1, A_2\) and latent response \(L,\) that
the average treatment effect for non-responders \(\Delta_{NR}\) is equal at stage 1 and 2, that the
average placebo effect \(\Delta_{\mathrm{placebo}}\) is also equal at stage
1 and 2, and that the residual standard deviation \(\sigma_\epsilon\) is
constant from baseline (\(t = 0\)) to stage 2 (\(t = 2\)) and equal in the
placebo and active treatment groups conditional on the latent placebo
response and previous outcome observation.

Our goal is to evaluate the inferential ability of each of the
estimators \(\hat{\theta}_1,\) \(\hat{\theta}_2,\) and
\(\hat{\theta}_w\) with \(w = 0.5\) to estimate the two ATEs of
interest: \(\Delta_{\mathrm{all}}\) and \(\Delta_{NR}.\) We are
interested in the impact of the size of the average placebo effect (APE
or \(\Delta_{\mathrm{placebo}}\)) and the variance in the outcome
\(\sigma^2_Y\) based on the assumption that these two quantities have a
strong influence on the classification ability of \(C_R(y_0, y_1).\) If
\(\sigma_\epsilon\) is large, then there is high variance in the outcomes which
will create significant overlap between placebo responders and
non-responders making classification difficult. If
\(\Delta_{\mathrm{placebo}}\) is large, then the gap between
\(\Delta_{\mathrm{all}}\) and \(\Delta_{NR}\) is larger and it should be
easier to distinguish between the placebo responders and non-responders.

The data for this simulation experiment are generated as follows:

\begin{itemize}
\item
  \(Y_0 \sim N(0, \sigma_\epsilon^2)\)
\item
  \(L \sim Binom(1, 0.5)\)
\item
   Treatment is assigned at random with 1/3 of participants receiving active treatment
  \(A_1 = 1\) and the remaining receiving placebo \(A_1 = 0.\)
\item
  \(Y_1 = Y_0 + \Delta_{NR} A_1 + \Delta_{\mathrm{placebo}} L(1 - A_1) + \epsilon_1\)
  where \(\epsilon_1 \overset{iid}{\sim} N(0,\sigma_\epsilon^2).\)
\item
  \(R = I(Y_1 - Y_0 \geq q_{p_r})\) where \(q_{p_r}\) is the \(p_r\)th
  quantile of the differences \(Y_1 - Y_0\) among the placebo group.

\item The classifier is 
    \(C_R(y_0, y_1) = I(y_1 - y_0 \geq q_{p_r}).\)
\item
  Phase 2 treatment is assigned randomly to the non-responders \(R_1 =0\) with
  half receiving active treatment (\(A_2 = 1\)) and half receiving placebo (\(A_2 = 0\)).
\item
  \(Y_2 = Y_1 + \Delta_{NR} A_2 + \Delta_{\mathrm{placebo}} L(1 - A_2) + \epsilon_2\)
  where \(\epsilon_2 \overset{iid}{\sim} N(0,\sigma_\epsilon^2).\)
\end{itemize}

Note that the true placebo response status is determined by the latent factor \(L\)
and not the observed responder indicator \(R\) and that we assume that
responder type \(L\) is the same at all time points.

The estimators are calculated as described in
Section~\ref{sec-prev-estimators}. The Stage 2 estimator
\(\hat{\theta}_2\) and weighted estimator \(\hat{\theta}_w\) are
calculated using two classifiers; first, thresholding as described above
is used to classify placebo response, and second, using an oracle
revealing the true placebo responder status \(L.\) The thresholding
approach can be interpreted as a conventional estimate that would be
reported in practice. The oracle approach can be interpreted as an
implausible best case scenario, where we have full information on the
responder status. The simulations support our theoretical results in
Section~\ref{sec-estimands} that even under these ideal conditions, the
weighted estimate that has been popularized throughout the SPCD
literature will always be biased for either well-defined quantity
\(\Delta_{\mathrm{all}}\) or \(\Delta_{NR}\) as long as there is a
placebo effect.

In the simulations, we hold weight \(w\) fixed at \(w = 0.5\) for
simplicity and to illustrate how the weighted estimator does not target
either \(\Delta_{NR}\) nor \(\Delta_{\mathrm{all}}.\) The response
threshold is fixed at the empirical 50th percentile of the placebo arm's
responses \(q_{0.5}.\)  \(\Delta_{\mathrm{all}}=0\) is also fixed as the
null to represent the scenario where the treatment provides no benefits
on average over the population mixture of placebo response and no
placebo response. The placebo response is varied from the null
\(\Delta_{\mathrm{placebo}} = 0\) to a relatively high placebo response
\(\Delta_{\mathrm{placebo}} = 1\) with uncertainty in \(Y\) varied from
\(\sigma_\epsilon = 0.001\) to \(\sigma_\epsilon = 10.\) 10,000 Monte Carlo samples were generated for each simulation setting. 

\begin{figure*}[htbp]

\centering{

\includegraphics[width=\textwidth]{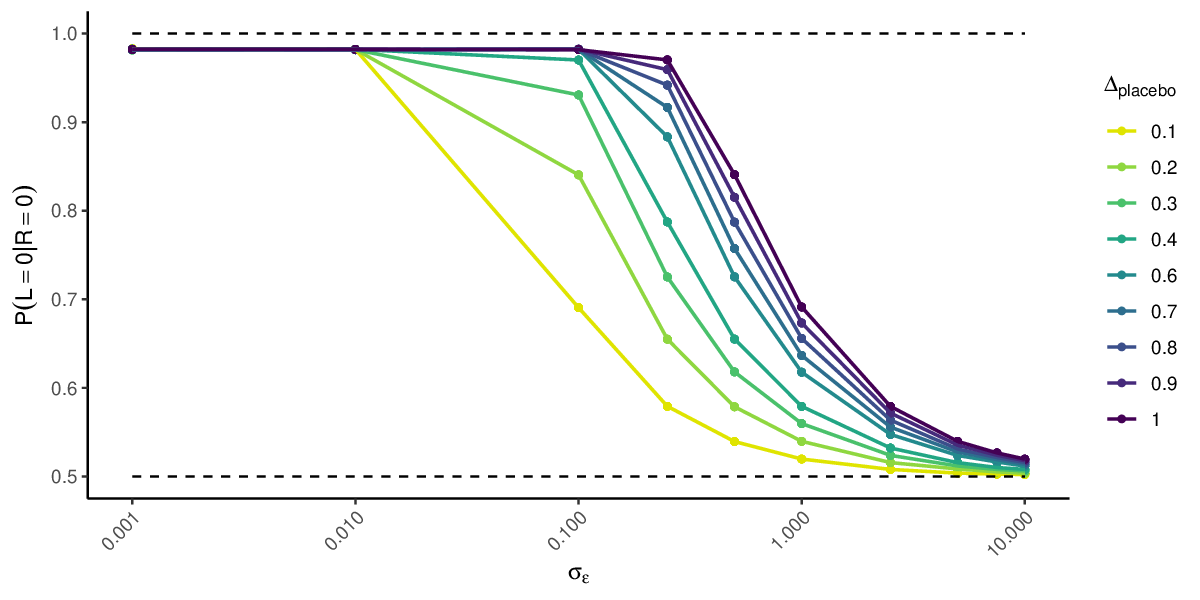}

}

\caption{\label{fig-sim-nonresp-npv}The average negative predictive values $P(L = 0 | R = 0)$ for the threshold classifier used in the simulation study.}

\end{figure*}%

In Figure~\ref{fig-sim-nonresp-npv}, we display the average negative predictive values (NPV; $P(L = 0 | R = 0)$) from the simulation study for the threshold classifier described in Section~\ref{sec-simulation}. The lowest possible NPV without knowledge of $L$ is 0.5 in our simulations since we set exactly half of participants to be placebo non-responders. A NPV of 0.5 then corresponds to assigning responder classifications at random (or equivalently calling everyone a placebo responder). A high NPV corresponds to correctly identifying the true placebo non-responders and only including those participants in the stage 2 re-randomization. 

We can see that when $\Delta_{\mathrm{placebo}}$ is small (the light lines and points) the NPV begins dropping towards the baseline 0.5 for lower values of $\sigma_\epsilon$ than for larger $\Delta_{\mathrm{placebo}}.$ This indicates that non-responder classification is easier when there is either a medium to large average placebo effect or small residual standard deviation relative to the average placebo effect. Both of these conditions result in clear separation between the modes of the mixture of placebo responders and non-responders.

\begin{figure*}[phtb]

\centering{

\includegraphics[width=\textwidth]{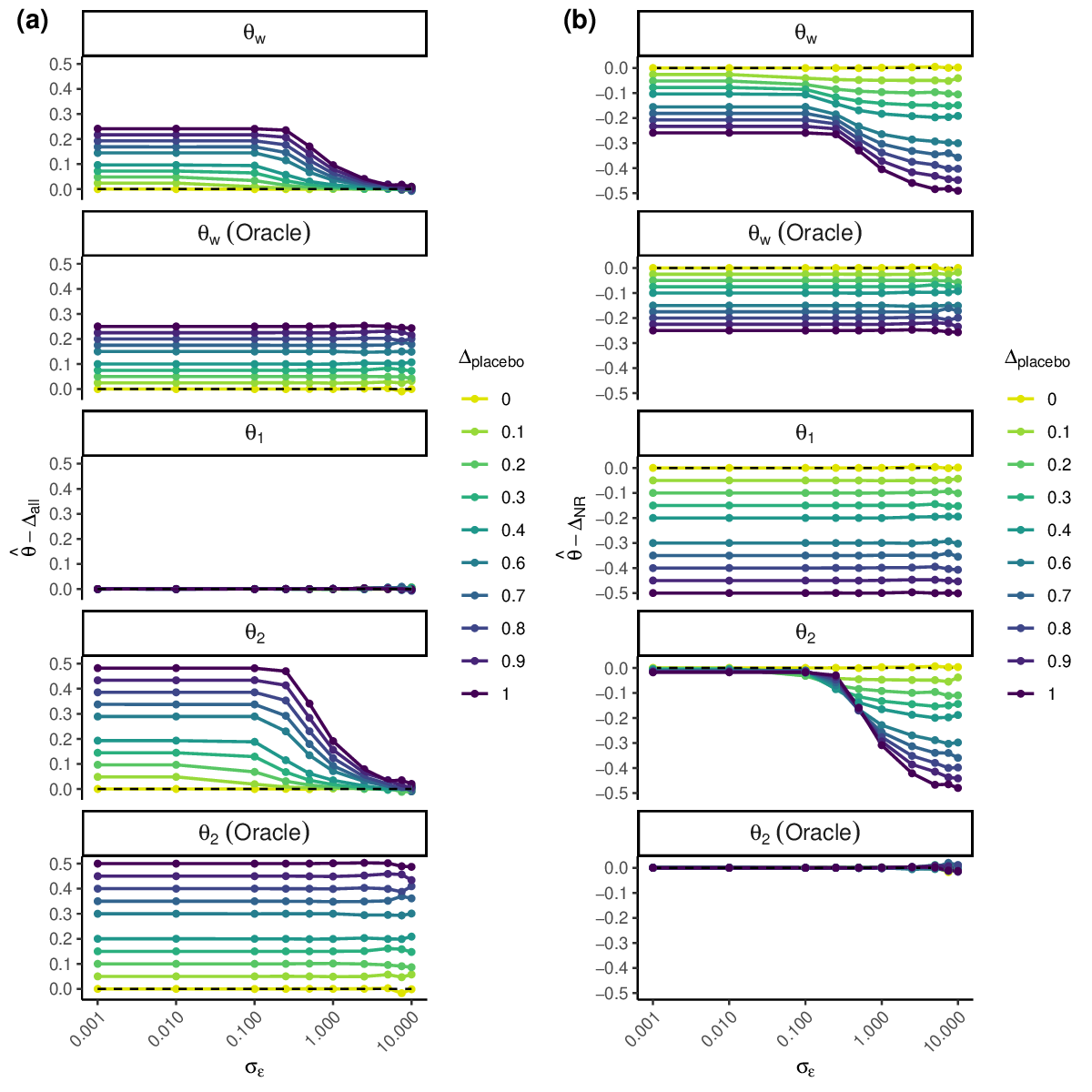}

}

\caption{\label{fig-sim-ate-sd-bias}Simulation under the null;
\(\Delta_{\mathrm{all}} = 0.\) Varying the average placebo effect (APE)
\(\Delta_{\mathrm{placebo}}\) and residual standard deviation
\(\sigma_\sigma\). The residual standard deviation is plotted along the x-axis. The y-axis in (a) is the estimated bias in estimating $\Delta_{\mathrm{all}}$ for each estimator. The y-axis in (b) is the estimated bias in estimating $\Delta_{NR}$ for each estimator. The lines and points are color-coded by the pre-set $\Delta_{\mathrm{placebo}}$.}

\end{figure*}

In Figure~\ref{fig-sim-ate-sd-bias}, the residual
standard deviation \(\sigma_\epsilon\) is displayed along the x-axis. The
y-axis displays the bias in estimating $\Delta_{\mathrm{all}}$ in Figure~\ref{fig-sim-ate-sd-bias}a and the bias in estimating $\Delta_{NR}$ in Figure~\ref{fig-sim-ate-sd-bias}b. The lines and points are color coded by the average placebo effect $\Delta_{\mathrm{placebo}}$ from 0 to 1 varied by increments of 0.1 with lower values in lighter colors and higher values in darker colors. The sub-panels correspond to the various estimators. ``Stage 1" (\(\hat{\theta}_1\)) measures the difference for the stage
1 treatment arms while ``Stage 2" (\(\hat{\theta}_2\)) measures the difference for
the stage 2 classified placebo non-responders and ``Weighted" (\(\hat{\theta}_w\)) is
the weighted average of these. The ``Stage 2 (Oracle)" and ``Weighted (Oracle)" are the stage 2 and weighted estimators applied to the true placebo non-responders. A horizontal black dashed line marks zero bias in each sub-panel. Deviations from the horizontal line at zero indicate bias in estimating the respective ATEs.

From Figure~\ref{fig-sim-ate-sd-bias}a we can see that
\(\hat{\theta}_1\) is consistently able to unbiasedly estimate
\(\Delta_{\mathrm{all}}.\) For moderately large
standard deviations (\(\sigma_\epsilon > 1\)) that exceed the average placebo
effect size, the non-oracle
estimators \(\hat{\theta}_2\) and \(\hat{\theta}_w\) tend towards \(\Delta_{\mathrm{all}}\) as $\sigma_\epsilon$ increases because the classifier
cannot correctly distinguish between responders and non-responders even
for large \(\Delta_{\mathrm{placebo}}\) as evidenced by Figure~\ref{fig-sim-nonresp-npv}. 

In Figure~\ref{fig-sim-ate-sd-bias}b, the stage 2 oracle estimator
\(\hat{\theta}_2^{\mathrm{oracle}}\) is unbiased for \(\Delta_{NR}\) as
long as \(\sigma_\epsilon\) is small (\textless{} 5), while the non-oracle
\(\hat{\theta}_2\) is only unbiased for \(\Delta_{NR}\) when \(\sigma_\epsilon < 1.\) The weighted estimator, \(\hat{\theta}_w,\) fails to estimate either \(\Delta_{NR}\)
or \(\Delta_{\mathrm{all}}\) when \(\Delta_{\mathrm{placebo}} > 0\) and
\(\sigma_\epsilon > 1,\) and \(\hat{\theta}_w^{\mathrm{oracle}}\) is always
biased for both \(\Delta_{NR}\) and \(\Delta_{\mathrm{all}}.\) 

Crucially, the weighted
estimator \(\hat{\theta}_w\) never estimates \(\Delta_{NR}\) and is
unbiased for either ATE only in the scenario where the SPCD provides no
benefit over a conventional parallel design because the classifier is
independent of true responder status \(P(L = 1 | R = 0) = P(L = 1)\).
The weighted oracle estimator \(\hat{\theta}_w^{\mathrm{oracle}}\) is
always biased for both ATE as long as \(\Delta_{\mathrm{placebo}}>0.\)
When \(\Delta_{\mathrm{placebo}} = 0,\) the estimands \(\Delta_{NR}\)
and \(\Delta_{\mathrm{all}}\) coincide and the estimators also coincide
in expectation, so they are all unbiased in this scenario.

\section{Example: ADAPT-A Benchmark}\label{sec-example}

As noted by Song et al.\cite{song2025}, the ADAPT-A trial\cite{fava_double-blind_2012} has served as a benchmark for methodological developments in the analysis of the SPCD\cite{tamura_examination_2007, chen2011, doros_repeated_2013, rybin2018, song2025}. Given its status as a benchmark data set, we defer extensive description and discussion of the data to its original publication\cite{fava_double-blind_2012} and other statistical methodology work that developed previous estimators\cite{chen2011, doros_repeated_2013, rybin2018, song2025}. In this section, we present our estimates of the conventional SPCD estimands for the ADAPT-A trial using our unadjusted estimators proposed in Section \ref{sec-estimands}. Alongside our estimates, we also present estimates using previously published methods including seemingly unrelated regression (SUR)\cite{tamura_examination_2007}, ordinary least squares (OLS)\cite{chen2011}, linear mixed models (MIX)\cite{doros_repeated_2013}, imputation within an EM algorithm (EMIMP)\cite{rybin2018}, and the unadjusted (UNADJ (RC)) and OLS (OLS (RC)) methods following the cross-validated placebo response reclassification proposed by Song et al.\cite{song2025} (where ``RC" stands for reclassification). Note that we do not include the estimators from Liu et al.\cite{liu2022} nor the G-computation, IPW, or doubly robust estimators from Song et al.\cite{song2025} as these estimators do not target the same quantities as the previously proposed conventional estimators. 

Interpreting the ADAPT-A trial within the latent-baseline framework of Song et al.\cite{song2025} would require positing an unobserved baseline placebo-response label and assuming that it is fully captured by baseline covariates, with the stage 1 responder classification acting as a simple surrogate. In ADAPT-A, however, the only operational definition of ``responder'' is a post-treatment function of the stage 1 outcome, and no baseline variables plausibly exhaust the drivers of placebo
response. Under a structural causal model that respects this actual SPCD
mechanism, the baseline responder label and its propensity are not identified from the data without strong,
unverifiable modeling assumptions; thus the latent-baseline framework is
not attainable for this trial. In other words, the assumptions required
by this latent-baseline formulation are implausibly strong for ADAPT-A
and reflect a clear mismatch with the SPCD design as actually
implemented.

\begin{table*}[h]
\begin{center}

\caption{\label{tbl-spcd-adapt-a-est}Estimates for the ADAPT-A trial benchmark data set from \cite{fava_double-blind_2012, song2025}. Our proposed unadjusted method is labeled ``UNADJ". The other methods and estimates were previously published: ``SUR" is seemingly unrelated regression from Tamura et al.\cite{tamura_examination_2007}, ``OLS" is ordinary least squares regression from Chen et al.\cite{chen2011}, ``MIX" is a linear mixed model from Doros et al.\cite{doros_repeated_2013}, ``EMIMP" is the EM algorithm-based estimation from Rybin et al.\cite{rybin2018}, and ``UNADJ (RC)" and ``OLS (RC)" are the unadjusted and ordinary least squares estimates using the placebo response reclassification step proposed by Song et al.\cite{song2025}.}
\centering
\begin{tabular*}{0.88\textwidth}{lrrrrrrr}
\toprule
\textbf{Method} & $\hat{\Delta}_{all}$ & $SE(\hat{\Delta}_{all})$ & $\hat{\Delta}_{NR}$ & $SE(\hat{\Delta}_{NR})$ & $\hat{\Delta}_{w=0.75}$ & $SE(\hat{\Delta}_{w=0.75})$ & \textbf{P-value} \\
\midrule
SUR & -0.526 & 0.955 & -2.228 & 1.125 & -0.952 & 0.769 & 0.216\\
OLS & -0.532 & 1.238 & -2.173 & 1.126 & -0.942 & 0.970 & 0.331\\
OLS (RC) & -0.532 & 1.164 & -1.914 & 1.053 & -0.878 & 0.912 & 0.336\\
MIX & -0.527 & 1.233 & -2.382 & 1.132 & -0.991 & 0.967 & 0.306\\
EMIMP & -0.447 & 1.146 & -2.353 & 1.493 & -0.924 & 0.937 & 0.324\\
\addlinespace
UNADJ & -0.447 & 1.195 & -2.197 & 1.112 & -0.885 & 0.938 & 0.346\\
UNADJ (RC) & -0.447 & 1.168 & -2.008 & 1.077 & -0.837 & 0.917 & 0.361\\
\bottomrule
\end{tabular*}
\end{center}
\end{table*}

Let $\hat{\theta}_t^{M}$ be the point estimate at stage $t$ with method $M$ from the set of methods $\{UNADJ, OLS, SUR, MIX, EMIMP\}$, and let $SE(\hat{\theta}_t^M)$ be the standard error for that point estimate. For the methods SUR, OLS, MIX, and EMIMP, the standard errors were calculated using closed form estimators. For the UNADJ, UNADJ (RC), and OLS$^*$ methods, the standard errors were approximated using a bootstrap with 1,000 replicates.

The estimates from each of these methods are displayed in Table \ref{tbl-spcd-adapt-a-est}. Note that the stage 1 estimates $\hat{\theta}_1^M$ from the unadjusted or OLS estimation methods are unaffected by the placebo response reclassification proposed by Song et al., i.e. $\hat{\theta}_1^{UNADJ} = \hat{\theta}_1^{UNADJ ~ (RC)} = -0.447$ and $\hat{\theta}_1^{OLS} = \hat{\theta}_1^{OLS ~ (RC)} = -0.532$. Additionally, the unadjusted stage 1 point estimate is also equivalent to the stage 1 point estimate with the EM algorithm, $\hat{\theta}_1^{UNADJ} = \hat{\theta}_1^{EMIMP} = -0.447$. However, all of the standard errors for these point estimates vary due to being calculated analytically or approximated via bootstrapping. The regression based methods (OLS, OLS$^*$, SUR, and MIX) all produced stage 1 point estimates of approximately -0.530. 

The range of the stage 2 estimates ($\hat{\theta}_2^{MIX} = -2.382$ to $\hat{\theta}_2^{OLS ~ (RC)} = -1.917$) is greater than the range of the stage 1 estimates ($\hat{\theta}_1^{OLS} = -0.532$ to $\hat{\theta}_1^{UNADJ} = -0.447$). The additional variation we see in stage 2 point estimates could be due to reducing the sample size at stage 2, or due to differences in how placebo response classification is addressed. In particular, the differences in placebo response classification due to the reclassification proposed by Song et al. results in the unadjusted and OLS estimates to be attenuated.

The large standard errors at both stages result in large standard errors for the SPCD estimator $\hat{\theta}_w^M$ for all of the methods as well. This results in no estimates being statistically significant at the $\alpha = 0.05$ level for the hypothesis test of $H_0: \Delta_{w=0.75}=0$ vs $H_1: \Delta_{w=0.75} \neq 0$. The p-values were calculated using a normal approximation $p = 2\left(1 - \Phi\left(|\hat{\Delta}_{w=0.75}|~/~SE(\hat{\Delta}_{w=0.75}\right)\right)$.

\section{Discussion}\label{sec-discussion}

In summary, we derived the estimands for each subset of participants
(all participants, placebo non-responders, and placebo responders) in
the SPCD while characterizing the placebo non-responder status as a
latent binary variable. The latent status is estimated by the classifier
\(C_R(y_0, y_1)\) using the baseline and stage 1 data for the placebo
arm. Further, we demonstrated that two of the conventional estimators for the
SPCD, \(\hat{\theta}_2\) (Equation \ref{eq-theta-2}) and \(\hat{\theta}_w,\) (Equation \ref{eq-theta-w}) do not target their
intended estimands while \(\hat{\theta}_1\) does unbiasedly estimate
\(\Delta_{\mathrm{all}}.\) Our analytical and simulation results
indicated that placebo non-response misclassification is a central cause
of the bias.

Specifically from Equation \ref{eq-exp-theta-2}, the conditions for which
\(E(\hat{\theta}_2) = \Delta_{NR}\), the desired expectation, are (i)
\(P(L = 1|R = 0) = 0,\) equivalently, \(R_i = 0 \Rightarrow L_i = 0\)
for all placebo non-responders (i.e.~the classifier only identifies true
non-responders as non-responders), or (ii)
\(\Delta_{\mathrm{placebo}}=0\). However, both of these conditions are
unverifiable. In the first condition (i) the probability of placebo
responder misclassification, \(P(L = 1 | R = 0)\), depends on the
unknown latent status \(L,\) so we cannot check the negative predictive
performance of \(C_R(y_1, y_0)\). The second condition (ii) requires
that the placebo has no effect. This condition is not testable, is not
typically a reasonable assumption, and defeats the purpose of using an
SPCD to estimate the placebo effect. Condition (i), the assumption of
perfect classification, is typically assumed implicitly in order to
obtain an unbiased estimator without acknowledging that it is impossible to check if it is even plausible\cite{rybin2018, chi2015, cui2019, rybin2015}.

Further, the weighted estimator, \(\hat{\theta}_w\), has been the primary
inferential tool in the SPCD literature\cite{chen2011, rybin2018, tamura2011, chi2015, rybin2015}, despite the
conceptual difficulty in interpreting its expectation given by Equation
\ref{eq-exp-theta-w}. Primarily, it is unclear how to interpret the
weighted average component \((wP(L=1) + (1-w)P(L=1|R=0))\) of the
expectation of \(\hat{\theta}_w\). This conceptual challenge can be
resolved under the above implausible conditions for unbiasedness for
\(\hat{\theta}_2\) or under the third condition (iii) that
\(P(L = 1|R = 0) = P(L = 1).\)

Under condition (iii), the expectation simplifies to
\(E(\hat{\theta}_w) = \Delta_{NR} - P(L = 1)\Delta_{\mathrm{placebo}} = \Delta_{\mathrm{all}}.\)
However, \(P(L = 1|R = 0) = P(L=1)\) occurs if \(R \perp L\),
\(P(L = 1) = 0\), \(\Delta_{\mathrm{placebo}}=0\), or \(w=1\) (see
Result \ref{res-est-w-all}). All four conditions defeat the purpose of
using the SPCD. The classified placebo responder status \(R\) being
independent of \(L\) implies that we do not gain any information by
performing the classification and stage 2 is essentially a replication
of stage 1 because true placebo responders are not excluded from stage
2. The second and third conditions
\((P(L = 1) = 0 \text{ or }\Delta_{\mathrm{placebo}} = 0)\) imply that no placebo responders are in the population to begin with. Lastly,
choosing \(w=1\) results in only using the stage 1 data which reduces
the SPCD to a traditional randomized two-arm trial.

If we are trying to estimate \(\Delta_{NR},\) then we should use
\(w = 0 \Rightarrow \hat{\theta}_w =\hat{\theta}_2\) and hope that
\(P(L = 1|R = 0)\) is small (see Result \ref{res-est-w-nr}). When
\(w = 0,\) the estimation only uses placebo arm stage 1 data to
determine \(C_R(y_0, y_1)\), and throws away the entire active treatment
arm where \(A_1 = 1.\) In order to accurately estimate \(\Delta_{NR}\)
with \(\hat{\theta}_2\), whichever (deterministic or stochastic)
placebo response classifier \(C_R(y_0, y_1)\) that is used must perfectly identify true responders \(L = 1.\) If that is not
the case then some true responders \(L = 1\) may be classified as
non-responders by \(R = 0\) creating a mixture at Stage 2.

Because the true latent statuses are unknown, there is uncertainty in the
classifications. In practice, we should require the inferential model or
estimators to account for the classification uncertainty in the variance
estimate of the stage 2 estimator. However, the classification
uncertainty is unaccounted for in previous definitions of the SPCD estimator\cite{chi2015,
cui2019, liu2022, song2025}. The previously shown
suboptimal control of Type I error by the weighted SPCD estimator
\(\hat{\theta}_w\) may be partially explained by the stage 2 estimator
failing to unbiasedly estimate \(\Delta_{NR}\) due to the
misclassification\cite{cui2019, liu2022}.

Of key importance to recognize is that the overall weighted estimator \(\hat{\theta}_w\) is particularly
problematic. Using any \(w \neq 0,1\) results in a biased weighted
estimator \(\hat{\theta}_w\) whose expectation (Equation
\ref{eq-exp-theta-w}) does not correspond to a quantity of clear
clinical interest. The placebo non-responder ATE is adjusted for placebo
response \(\Delta_{\mathrm{placebo}}\) by some weighted average of the
proportion of placebo responders in the population and the proportion of
responders that are misclassified as non-responders. It does not seem
likely that this quantity would be of clinical interest or importance
given that it is a function of the statistical analysis through the
classifier \(C_R(y_0, y_1).\)

Moreover, it is still unclear how to interpret what the combination of
the SPCD stage 1 and stage 2 estimators measure when
\(\Delta_{\mathrm{all}}\) and \(\Delta_{NR}\) are not equal, even in the
scenarios where we can unbiasedly estimate \(\Delta_{\mathrm{all}}\) or
\(\Delta_{NR}\). We have not identified an explanation and description
of this estimator, grounded in clinical meaning, of the weighted average
of a marginal and conditional probability for the placebo non-response
mean.

Our results are related to prior work\cite{liu2022, song2025}, but highlight the importance of the
accuracy of classification and provide a quantification of the bias of
estimates due to misclassification. Where the previous work\cite{liu2022} uses a counterfactual variable to define
the responder status so that both drug and placebo
responders and non-responders are defined resulting in four strata, we
define only placebo response through our non-counterfactual, latent
status \(L\) which assumes that everyone would be a drug responder, similar to the approach from Song et al.\cite{song2025} This
in effect collapses Liu et al.'s never-responder and drug-only responder strata\cite{liu2022}, while the
always-responder strata correspond to our placebo responders and active treatment
participants.

Additionally, Liu et al.\cite{liu2022} focus on the
estimation of the individual strata effects at each time point, and the
effects for never-responders and drug-only responders by taking a
weighted average of their respective estimates at each stage. Their
results assume that classification is accurate following the EM
algorithm's maximum likelihood estimation and sampling of responder
status. Our results provide insight of the impact of misclassification on the estimates.

While Song et al.\cite{song2025} advocate for using their newly proposed estimands and estimators with a novel approach to classification, they acknowledge that misclassification will result in bias even for their estimators, and that misclassification is unavoidable. Additionally, despite acknowledging the difficulty in interpreting the SPCD estimator $\hat{\Delta}_w$ because it mixes study populations, Song et al. advocates for adopting the $\Delta_{Joint}$ estimand which also mixes study populations in an uninterpretable manner (participants in the first stage vs participants in the second stage).

Given the consensus on the importance of accurate placebo response classification in the inferential procedure for SPCD analysis, the intractability of placebo response misclassification, and the demonstrated amount of bias due to misclassification shown in Section~\ref{sec-app-theta-hat-deriv} as well as in prior work\cite{liu2022, song2025}, we would advocate for using alternative approaches to account for placebo response in clinical trials such as optimal partitioning via finite mixture models\cite{tarpey_optimal_2010} or latent regression analysis where placebo response is reconsidered as a continuous latent predictor\cite{tarpey_latent_2010}.

Beyond the scope of this investigation, it may be
better to model placebo response as a continuous latent variable rather
than a binary indicator since it is plausible that everyone can exhibit some degree of placebo response along a continuum. While the binary responder status is a
simplifying assumption that ties estimation to the well-known finite mixture model, it may not be a
reasonable assumption for many of the applications in which SPCD are
deployed, such as in the study of interventions for pain or psychiatric
conditions.


\subsubsection*{Funding Statement}

This research was supported in part by Grant Number 5U24NS113844-0 from the National Institute of Neurologic Disorders and Stroke and by NYU CTSA grant (UL1TR001445) from the National Center for Advancing Translational Sciences, National Institutes of Health. 

This research is based upon work supported by the National Science Foundation under Grant No 2306556 and National Institute of Health Grant No 1R01AI197146-01.

\subsubsection*{Financial disclosure}

None reported.

\subsubsection*{Conflict of interest}

The authors declare no potential conflict of interests.

\printbibliography

\newpage

\appendix

\subsection*{Estimand Derivations}\label{sec-app-est-deriv}

We provide identification results for each of the stage-1 estimands
under the structural causal model described in
Section~\ref{sec-background}, assuming Assumptions
\ref{asm-cons_sutva}--\ref{asm-mixture}.

\begin{result}[Average Placebo Effect] \label{res-ape}
The average effect due to placebo is
$$
\Delta_{\text{placebo}} = E(Y_1^{a_1=0,l=1}) - E(Y_1^{a_1=0,l=0}).
$$
\end{result}
\begin{proof}

First, we establish the identification of the expectation of the counterfactuals from the observed data. For $l \in \{0,1\},$

\begin{align*}
E(Y_1^{a_1=0,l})
&= E(Y_1^{a_1=0,l} \mid A_1=0)
      &&\text{by Assumption~\ref{asm-exch}} \\
&= E(Y_1^{a_1=0,l} \mid A_1=0, L=l)
      &&\text{by Assumption~\ref{asm-latent}} \\
&= E(Y_1 \mid A_1=0, L=l)
      &&\text{by Assumptions~\ref{asm-cons_sutva},\,\ref{asm-positivity}} \\
&\text{identified from } f_{Y_1 \mid A_1=0}(y)
      &&\text{by Assumption~\ref{asm-mixture}}
\end{align*}

Then,
\begin{align*}
E(Y_1^{a_1 = 0, l=1} &- Y_0) - E(Y_1^{a_1 = 0, l=0} - Y_0) = \left[E(Y_1^{a_1 = 0, l=1}) - E(Y_0)\right] - \left[E(Y_1^{a_1 = 0, l=0}) - E(Y_0)\right] \\
&= E(Y_1^{a_1 = 0, l=1}) - E(Y_1^{a_1 = 0, l = 0}) \\
&= \Delta_{\mathrm{placebo}}. \label{eq-proof-ate-placebo} \numberthis 
\end{align*}

Note that $E(Y^{a_1 = 0, l=1}) = \Delta_{\mathrm{placebo}} + E(Y^{a_1 = 0, l=0}).$

\end{proof}

\begin{result}[Average Treatment Effect for Placebo Non-Responders] \label{res-ate-nr}
The average treatment effect for true placebo non-responders is
$$
\Delta_{\text{NR}} = E[Y_1^{a_1=1,l=0}] - E[Y_1^{a_1=0,l=0}].
$$
\end{result}
\begin{proof}
Again, we begin by identifying the counterfactual's expectations from the data.

\begin{align*}
E(Y_1^{1,0})
&= E(Y_1^{1}) &&\text{by Assumption~\ref{asm-homogeneity}} \\
&= E(Y_1 \mid A_1 = 1)
      &&\text{by Assumptions~\ref{asm-cons_sutva},\,\ref{asm-exch},\,\ref{asm-positivity}} \\[4pt]
E(Y_1^{0,0})
&= E(Y_1 \mid A_1=0, L=0)
      &&\text{by Assumptions~\ref{asm-cons_sutva},\,\ref{asm-exch},\,\ref{asm-positivity},\,\ref{asm-latent}} \\
&\text{identified from } f_{Y_1 \mid A_1=0}(y)
      &&\text{by Assumption~\ref{asm-mixture}}
\end{align*}

Then,
\begin{align*}
E(Y_1^{a_1 = 1} - Y_0) &- E(Y_1^{a_1 = 0, l=0} - Y_0) = \left[E(Y_1^{a_1 = 1}) - E(Y_0)\right] - \left[E(Y_1^{a_1 = 0, l = 0}) - E(Y_0)\right] \\
&= E(Y_1^{a_1=1}) - E(Y_1^{a_1=0, l=0}) \\
&= \Delta_{NR}. \label{eq-proof-ate-nr} \numberthis 
\end{align*}

Note that $E(Y_1^{a_1=1}) - E(Y_1^{a_1=0, l=0}) = \Delta_{NR}.$

\end{proof}

\begin{result}[Average Treatment Effect for Placebo Responders] \label{res-ate-pr}
The average treatment effect for true placebo responders is
$$
\Delta_{\text{PR}} = E(Y_1^{a_1=1,l=1}) - E(Y_1^{a_1=0,l=1}).
$$
We can also summarize the ATE for placebo responders as the quantity $\Delta_{PR} = \Delta_{NR} - \Delta_{placebo}.$
\end{result}
\begin{proof}
\begin{align*}
E(Y_1^{a_1 = 1,l=1})
&= E(Y_1^{a_1 = 1,l=0}) = E(Y_1 \mid A_1 = 1)
      &&\text{by Assumptions~\ref{asm-homogeneity},\,\ref{asm-cons_sutva},\,\ref{asm-exch},\,\ref{asm-positivity}} \\[4pt]
E(Y_1^{a_1 = 0,l=1})
&= E(Y_1 \mid A_1 = 0, L = 1)
      &&\text{by Assumptions~\ref{asm-cons_sutva},\,\ref{asm-exch},\,\ref{asm-positivity},\,\ref{asm-latent}} \\
&\text{identified from } f_{Y_1 \mid A_1=0}(y)
      &&\text{by Assumption~\ref{asm-mixture}}
\end{align*}

\begin{align*}
\Delta_{PR} &= E(Y_1^{a_1 = 1}) - E(Y_1^{a_1 = 0, l=1}) \\
&= E(Y_1^{a_1 = 1}) - \left[E(Y_1^{a_1 = 0, l=0}) + \Delta_{\mathrm{placebo}}\right] \\
\Rightarrow \Delta_{\text{PR}}
&= \Delta_{\text{NR}} - \Delta_{\text{placebo}}
      &&\text{by algebra under Assumption~\ref{asm-homogeneity}}. \label{eq-proof-ate-r} \numberthis 
\end{align*}
\end{proof}

\begin{result}[Average Treatment Effect for All Participants - Difference] \label{res-ate-all-1}
The average treatment effect for all participants can be represented as a simple difference
$$
\Delta_{\text{all}} = E(Y_1^{a_1=1}) - E(Y_1^{a_1=0}).
$$
\end{result}
\begin{proof}
\begin{align*}
E(Y_1^{a})
&= E(Y_1 \mid A_1 = a)
      &&\text{by Assumptions~\ref{asm-cons_sutva},\,\ref{asm-exch},\,\ref{asm-positivity}} \\[4pt]
\Rightarrow
\Delta_{\text{all}}
&= E(Y_1 \mid A_1=1) - E(Y_1 \mid A_1=0).
\end{align*}
\end{proof}
\begin{result}[Average Treatment Effect for All Participants - Mixture] \label{res-ate-all-2}
The ATE for all participants can also be represented as a mixture

$$
\Delta_{\text{all}}
  \;=\; (1 - P(L=1))\Delta_{\text{NR}} + P(L=1)\Delta_{\text{R}}
  \;=\; \Delta_{\text{NR}} - P(L=1)\Delta_{\text{placebo}}.
$$
\end{result}
\begin{proof}
Next, the ATE for the full population is presented in mixture form. $P(L=1)$ is identified from $f_{Y_1 \mid A_1 = 0}(y)$ by Assumptions \ref{asm-positivity} and \ref{asm-mixture}. $\Delta_{\text{NR}},\;\Delta_{\text{PR}},\;\Delta_{\text{placebo}}$ are identified as in Results \ref{res-ape}-\ref{res-ate-pr}. Then from algebra,

\begin{align*}
E(Y_1^{a_1 = 1}) - E(Y_1^{a_1 = 0, l}) &= E(Y_1^{a_1 = 1}) - E\left(P(L=1) Y_1^{a_1 = 0, l=1} + (1 - P(L=1))Y_1^{a_1 = 0, l = 0}\right) \\
&= E(Y_1^{a_1 = 1}) - \left[P(L=1) E(Y_1^{a_1 = 0, l=1}) + (1 - P(L=1)) E(Y_1^{a_1 = 0, l = 0})\right] \\
&= E(Y_1^{a_1 = 1}) - P(L = 1) (\Delta_{\mathrm{placebo}} + E(Y_1^{a_1 = 0, l=0})) - (1 - P(L=1)) E(Y_1^{a_1 = 0, l=0}) \\
&= E(Y_1^{a_1=1}) - E(Y_1^{a_1 = 0, l=0}) - P(L=1) \Delta_{\mathrm{placebo}} \\
&= \Delta_{NR} - P(L=1) \Delta_{\mathrm{placebo}}. \label{eq-proof-ate-all} \numberthis 
\end{align*}
\end{proof}

\subsection*{Estimator Derivations}\label{sec-app-theta-hat-deriv}

\begin{result}[Expectation of Stage 1 Estimator] \label{res-est-1}
$$
E(\hat{\theta}_1) = \Delta_{\mathrm{all}}.
$$
\end{result}
\begin{proof}
We make Assumptions \ref{asm-cons_sutva}-\ref{asm-stationarity}. Then,
\begin{align*}
    E(\hat{\theta}_1) &= E\left[\frac{1}{n_A} \sum_{i=1}^{n} (Y_{i, 1} - Y_{i,0})I(A_{i,1} = 1) - \frac{1}{n_P} \sum_{i=1}^{n} (Y_{i, 1} - Y_{i,0})I(A_{i,1} = 0)\right] \\
    &= \frac{1}{n_A} \sum_{i=1}^{n} E(Y_{i, 1} - Y_{i,0} | A_{i,1} = 1) - \frac{1}{n_P} \sum_{i=1}^{n} E(Y_{i, 1} - Y_{i,0} | A_{i,1} = 0) \\
    &= \frac{1}{n_A} \sum_{i=1}^{n_A} E(Y_{i, 1}^{a_1 = 1} - Y_{i,0}) - \frac{1}{n_P} \sum_{i=1}^{n_P} E(Y_{i, 1}^{a_1 = 0, l} - Y_{i,0}) \\
    &= E(Y_{1}^{a_1 = 1} - Y_{0}) - E(Y_{1}^{a_1 = 0, l} - Y_{0}) \\
    &= E(Y_{1}^{a_1 = 1}) - E(Y_{1}^{a_1 = 0, l}) \\
    &= \Delta_{\mathrm{all}}. \label{eq-proof-exp-theta-1} \numberthis 
\end{align*}
\end{proof}

\begin{result}[Expectation of the Stage 2 Estimator] \label{res-est-2}
$$
E(\hat{\theta}_2) = \Delta_{NR} - q_1 \Delta_{\mathrm{placebo}}.
$$

$E(\hat{\theta}_2) = \Delta_{NR}$ if $q_1 = P(L=1|R=0) = 0,$ i.e. if no true placebo responders $L=1$ are classified as placebo non-responders $R=0.$
\end{result}
\begin{proof}
Let $P(L = 0, R=0) = p_{00}$ and $P(L = 1, R = 0) = p_{10}$ so that $q_l = P(L = l | R = 0) = p_{l0} / (p_{00} + p_{10}).$ To calculate these probabilities, we would need to make a distributional assumption about $Y_t.$ 

We also define

\begin{align*}
E(Y_2^{a_1, a_2, l} | R = 0) &= P(L = 0|R = 0)E(Y_2^{a_1, a_2, l=0}|R=0) \\
& ~~~~~~~~ + P(L=1|R=0)E(Y_2^{a_1, a_2, l=1}|R=0) \\
    &= q_0 E(Y_2^{a_1, a_2, l=0}|R=0) + q_1 E(Y_2^{a_1, a_2, l=1}|R=0) \label{eq-proof-exp-theta-2} \numberthis 
\end{align*}

Again we make Assumptions \ref{asm-cons_sutva}-\ref{asm-stationarity} (modified accordingly for Stage 2), so that the placebo non-responder average treatment effect is the same as from Stage 1, i.e.

$$
E(Y_{2}^{a_1 = 0, a_2 = 1, l = 0}) - E(Y_{2}^{a_1 = 0, a_2 = 0, l = 0}) = E(Y_1^{a_1 = 1}) - E(Y_1^{a_0 = 0, l = 0})
= \Delta_{NR}.
$$

Note that given $R$, the sample sizes for each stage 2 arm are fixed $n_{PA}$ and $n_{PP}$ since the classified responder status determines the samples size of each arm. Then we get

\begin{align*}
E(\hat{\theta}_2) = &E\left[\frac{1}{n_{PA}} \sum_{i=1}^{n} (Y_{i, 2} - Y_{i,1})I(A_{i,1} = 0, R_i = 0, A_{i,2} = 1) | R\right] \\
&~~~~ - E\left[\frac{1}{n_{PP}} \sum_{i=1}^{n} (Y_{i, 2} - Y_{i,1})I(A_{i,1} = 0, R_i = 0, A_{i,2} = 0) | R\right] \\
&= \frac{1}{n_{PA}} \sum_{i=1}^{n_{PA}} E\left[Y_{i, 2} - Y_{i,1}|A_{i,1} = 0, R_i = 0, A_{i,2} = 1, \right] \\
&~~~~ - \frac{1}{n_{PP}} \sum_{i=1}^{n_{PP}} E\left[Y_{i, 2} - Y_{i,1} | A_{i,1} = 0, R_i = 0, A_{i,2} = 0\right] \\
&= \frac{n_{PA}}{n_{PA}} E\left[Y_{2}^{a_1=0, a_2 = 1, l} - Y_{1}^{a_1=0, l}|R = 0\right] \\
&~~~~ - \frac{n_{PP}}{n_{PP}} E\left[Y_{2}^{a_1=0, a_2 = 0, l} - Y_{1}^{a_1=0,  l} | R = 0\right] \\
&= E\left[Y_{2}^{a_1=0, a_2 = 1, l} - Y_{1}^{a_1=0, l}|R = 0\right] \\
&~~~~ - E\left[Y_{2}^{a_1=0, a_2 = 0, l} - Y_{1}^{a_1=0,  l} | R = 0\right] \\
&= E\left[Y_{2}^{a_1=0, a_2 = 1, l}|R = 0\right] - E\left[Y_{2}^{a_1=0, a_2 = 0, l} | R = 0 \right] \\
&= E\left[Y_{2}^{a_1=0, a_2 = 1, l}|R = 0\right] - E\left[Y_{2}^{a_1=0, a_2 = 0, l} | R = 0 \right] \\
&= \left[q_0 E(Y_2^{a_1=0, a_2=1, l=0}|R=0) + q_1 E(Y_2^{a_1=0, a_2=1, l=1}|R=0)\right] \\
&~~~~ - \left[q_0 E(Y_2^{a_1=0, a_2=0, l=0}|R=0) + q_1 E(Y_2^{a_1=0, a_2=0, l=1}|R=0)\right] \\
&= q_0\left[E(Y_2^{a_1=0, a_2=1, l=0}|R=0) - E(Y_2^{a_1=0, a_2=0, l=0}|R=0)\right] \\
&~~~~ + q_1 \left[ E(Y_2^{a_1=0, a_2=1, l=1}|R=0) - E(Y_2^{a_1=0, a_2=0, l=1}|R=0)\right] \\
&= q_0 \Delta_{NR} + q_1 \Delta_{PR}, ~~~~ \mbox{where } \Delta_{PR} = \Delta_{NR} - \Delta_{\mathrm{placebo}} \mbox{ is the placebo response.} \\
&= q_0 \Delta_{NR} + q_1 (\Delta_{NR} - \Delta_{\mathrm{placebo}}) \\
&= \Delta_{NR} (q_0 + q_1) - q_1 \Delta_{\mathrm{placebo}} \\
\Rightarrow E(\hat{\theta}_2) &= \Delta_{NR} - q_1 \Delta_{\mathrm{placebo}} \neq \Delta_{\mathrm{all}}. \label{eq-proof-exp-theta-2-2} \numberthis 
\end{align*}

\end{proof}

\begin{result}[Expectation of the Weighted SPCD Estimator] \label{res-est-w}
$$
E(\hat{\theta}_w) =\Delta_{NR} - \Delta_{\mathrm{placebo}}(wP(L=1) + (1-w)P(L=1|R=0)).
$$
\end{result}
\begin{proof}
\begin{align*}
E(\hat{\theta}_w) &= wE(\hat{\theta}_{1,\mathrm{all}}) + (1 - w)E(\hat{\theta}_2) \\
&= w\Delta_{\mathrm{all}} + (1 - w) \left[\Delta_{NR} - q_1 \Delta_{\mathrm{placebo}}\right] \\
&= w(\Delta_{NR} - p_L \Delta_{\mathrm{placebo}}) + (1-w)(\Delta_{NR} - q_1 \Delta_{\mathrm{placebo}}) \\
&= \Delta_{NR}(w + (1-w)) - \Delta_{\mathrm{placebo}} (wp_L + (1-w)q_1) \\
&= \Delta_{NR} - \Delta_{\mathrm{placebo}}(wP(L=1) + (1-w)P(L=1|R=0)).  \label{eq-proof-exp-theta-w} \numberthis 
\end{align*}
\end{proof}
\begin{result}[Weighted SPCD Estimator Targets $\Delta_{\mathrm{all}}$] \label{res-est-w-all}

$\hat{\theta}_w$ is unbiased for $\Delta_{\mathrm{all}}$ if any of the following hold:
\begin{enumerate}
\item  $P(L=1) = 0.$
\item  $\Delta_{\mathrm{placebo}}=0.$ 
\item  $R \perp\!\!\!\perp L.$ 
\item  We set $w = 1.$
\end{enumerate}
\end{result}
\begin{proof}

\begin{enumerate}[font=\itshape]
\item Suppose  $P(L=1)=0$.

By Result \ref{res-ate-all-2}, $\Delta_{\mathrm{all}} = \Delta_{NR}.$ Then,

\begin{align*}
E(\hat{\theta}_w) &=  \Delta_{NR} - \Delta_{\mathrm{placebo}}(wP(L=1) + (1-w)P(L=1|R=0)) \\
 &= \Delta_{NR} - \Delta_{\mathrm{placebo}} (w0 + (1-w)0) ~~~~ \mbox{since } P(L=1)=0 \Rightarrow P(L=1|R=0) = 0 \\
 &= \Delta_{NR} = \Delta_{\mathrm{all}}. \label{eq-proof-exp-theta-w-2-1} \numberthis 
\end{align*}

\item Suppose $\Delta_{\mathrm{placebo}}=0$.

By Result \ref{res-ate-all-2}, $\Delta_{\mathrm{all}} = \Delta_{NR}$. Then,
\begin{align*}
E(\hat{\theta}_w) &= \Delta_{NR} - 0(wP(L=1) + (1-w)P(L=1|R=0)) \\
&= \Delta_{NR} = \Delta_{\mathrm{all}}. \label{eq-proof-exp-theta-w-2-2} \numberthis 
\end{align*}

\item Suppose $R \perp\!\!\!\perp L$. 

Then $P(L=1|R=0) = P(L=1) = p_{L},$ and, 

\begin{align*}
E(\hat{\theta}_w) &=  \Delta_{NR} - \Delta_{\mathrm{placebo}}(wP(L=1) + (1-w)P(L=1|R=0)) \\
 &= \Delta_{NR} - \Delta_{\mathrm{placebo}} (wp_L + (1-w)p_L) \\
 &= \Delta_{NR} - p_L \Delta_{\mathrm{placebo}} \\
 &= \Delta_{\mathrm{all}}. \label{eq-proof-exp-theta-w-3} \numberthis 
\end{align*}

Note $R \perp\!\!\!\perp L$ is trivially true if $C_R$ assigns every participant $R_i = 0.$

\item 

Suppose we set $w = 1$. 

Then $\hat{\theta}_w = \hat{\theta}_1.$ See Result \ref{res-est-1} for the proof $\hat{\theta}_1$ is unbiased for $\Delta_{\mathrm{all}}.$
\end{enumerate}

\end{proof}
\begin{result}[Weighted SPCD Estimator Targets $\Delta_{NR}$] \label{res-est-w-nr}
$\hat{\theta}_w$ is unbiased for $\Delta_{NR}$ if $w = 0 \Rightarrow \hat{\theta}_w = \hat{\theta}_2$ and $\hat{\theta}_2$ is unbiased for $\Delta_{NR}.$
\end{result}
\begin{proof}
See Result \ref{res-est-2} for proof of when $\hat{\theta}_2$ is unbiased for $\Delta_{NR}.$
\end{proof}

\end{document}

%% file: spcd_diagram.tex
%
%
\begin{tikzpicture}[
    >=stealth,
    node distance=15mm,
    splitnode/.style={
        matrix of nodes,
        nodes={draw=none, align=center, text height=1.5ex, text depth=.25ex},
        row sep=0pt,
        column sep=0pt,
        rounded corners,
        draw,
        fill=white,
        inner sep=4pt,
        minimum width=5mm
    },
    stagebox/.style={
        rounded corners,
        draw,
        fill=gray!20,
        inner sep=7pt,
        minimum width=20mm,
        minimum height=60mm,
        fit=#1,
        opacity=0.33
    }
]

\coordinate (A) at (0,0);
\matrix (A) [splitnode, left=of A] {$N$ \\ participants \\};
\matrix (B) [splitnode, right=of A, yshift=15mm] {\textbf{Placebo} \\ $L = 0,1$ \\ $A_1 = 0$ \\};
\matrix (C) [splitnode, right=of A, yshift=-15mm] {\textbf{Active} \\ $L = 0,1$ \\ $A_1 = 1$ \\};
\matrix (D) [splitnode, right=of B, yshift=20mm] {\textbf{Classified} \\ \textbf{Non-responders} \\ $L = 0,1$ \\$R = 0$ \\$A_1 = 0$\\};
\matrix (E) [splitnode, right=of B, yshift=-20mm,xshift=3.5mm] {\textbf{Classified} \\ \textbf{Responders}\\ $L = 0,1$\\ $R = 1$ \\$A_1 = 0$ \\};
\matrix (F) [splitnode, right=of D, yshift=20mm] {\textbf{Placebo} \\ $L = 0,1$ \\$R = 0$ \\$A_1 = 0$\\ $A_2 = 0$\\};
\matrix (G) [splitnode, right=of D, yshift=-20mm] {\textbf{Active} \\ $L = 0,1$ \\$R = 0$ \\$A_1 = 0$\\ $A_2=1$\\};

\coordinate (labelbase) at ($(A.south)!-3cm!(C.south)$);
\coordinate (labeltop) at ($(A.north)!-7cm!(C.north)$);

\node[draw=none, below=3cm of labelbase -| B](label1) {};
\node[draw=none, below=3cm of labelbase -| D](label2) {};
\node[draw=none, below=3cm of labelbase -| F](label3) {};

\node[draw=none, above=4.25cm of labeltop -| B](label4) {\textbf{Stage 1}};
\node[draw=none, above=4.25cm of labeltop -| D](label5a) {\textbf{Placebo Response}};
\node[draw=none, above=3.75cm of labeltop -| D](label5b) {\textbf{Classification}};
\node[draw=none, above=4.25cm of labeltop -| F](label6) {\textbf{Stage 2}};

%

\begin{pgfonlayer}{background}
    \node[stagebox=(B)(C)(label1.south)(label4)] (Stage1) {};
    \node[stagebox=(D)(E)(label2.south)(label5a)] (Classification) {};
    \node[stagebox=(F)(G)(label3.south)(label6)] (Stage2) {};
\end{pgfonlayer}

\draw[->] (A.east) -- (B.west) node[midway, sloped, above, draw=none] {$A_1 = 0$};
\draw[->] (A.east) -- (C.west) node[midway, sloped, below, draw=none] {$A_1 = 1$};
\draw[->] (B.east) -- (D.west) node[midway, sloped, above, draw=none] {$R = 0$};
\draw[->] (B.east) -- (E.west) node[midway, sloped, below, draw=none] {$R = 1$};
\draw[->] (D.east) -- (F.west) node[midway, sloped, above, draw=none] {$A_2 = 0$};
\draw[->] (D.east) -- (G.west) node[midway, sloped, below, draw=none] {$A_2 = 1$};

\end{tikzpicture}
%